\documentclass[12pt,letter]{article}

\usepackage{latexsym,amsmath,amssymb,amsthm,amsfonts,bbm,graphicx,natbib,breqn,epsfig,bm,url,thmtools,xcolor}
\makeatletter 
\let\@fnsymbol\@arabic
\makeatother

\usepackage{float}   
\usepackage{booktabs} 
\usepackage{graphicx} 
\usepackage[margin=1cm]{caption} 
\usepackage{mathtools}
\usepackage{subfigure}

\floatstyle{plain}
\newfloat{Algorithm}{thp}{lop}
\floatname{Algorithm}{Algorithm}

\def\1{1\!{\rm l}}

\declaretheoremstyle[
notefont=\bfseries, notebraces={}{},
bodyfont=\normalfont,
postheadspace=0.5em,
numbered=yes,
]{mystyle}

\theoremstyle{definition}

\newcommand{\dt}{\text{d}}

\usepackage{subfigure}
\usepackage{algorithm}
\usepackage{algorithmic}

\usepackage{bigfoot}

\DeclareNewFootnote{AAffil}[arabic]
\DeclareNewFootnote{ANote}[fnsymbol]

\usepackage{etoolbox}
\makeatletter
\patchcmd\maketitle{\def\@makefnmark{\rlap{\@textsuperscript{\normalfont\@thefnmark}}}}{}{}{}
\makeatother

\makeatletter
\def\thanksAAffil#1{
  \footnotemarkAAffil\protected@xdef\@thanks{\@thanks%
        \protect\footnotetextAAffil[\the \c@footnoteAAffil]{#1}}%
}
\def\thanksANote#1{%
  \footnotemarkANote%
  \protected@xdef\@thanks{\@thanks%
        \protect\footnotetextANote[\the \c@footnoteANote]{#1}}%
}
\makeatother

\title{Modularized Bayesian analyses and cutting feedback in likelihood-free inference}

\author{
  Atlanta Chakraborty%
  \thanksAAffil{Institute for Operations Research and Analytics, National University of Singapore, Singapore 119077}%
  , %
  David J. Nott%
  \footnotemarkAAffil[1]$^{,}$\thanksAAffil{Department of Statistics and Data Science, National University of Singapore, Singapore 117546}%
  , %
  Christopher Drovandi%
  \thanksAAffil{School of Mathematical Sciences and Centre for Data Science, Queensland University of Technology, Brisbane 4000 Australia}%
  , \\%
  David T. Frazier%
  \thanksAAffil{Department of Econometrics and Business Statistics, Monash University, Clayton VIC 3800, Australia}%
  \text{ and} %
  Scott A. Sisson%
  \thanksAAffil{UNSW Data Science Hub \& School of Mathematics and Statistics, UNSW Sydney, Australia}%
}
\date{\empty}

\begin{document}

  
\maketitle

\vspace{-0.3in}
\begin{abstract}
\noindent There has been much recent interest in modifying Bayesian inference for misspecified
models so that it is useful for specific purposes.
One popular modified Bayesian inference method is ``cutting feedback''
which can be used when the model consists of a number of coupled modules,
with only some of the modules being misspecified.
Cutting feedback methods 
represent the full posterior distribution in terms of conditional and sequential
components, and then modify some terms in 
such a representation based on the modular structure for specification or computation
of a modified posterior distribution.  
The main goal of this is to 
avoid contamination of inferences for parameters of interest by misspecified modules.  
Computation for cut posterior distributions is challenging, and here we consider cutting feedback for
likelihood-free inference based on Gaussian mixture approximations to the joint distribution of parameters and 
data summary statistics.  
We exploit the fact that marginal and conditional distributions of a Gaussian mixture are Gaussian mixtures 
to give explicit approximations to marginal or conditional posterior distributions so that we can easily approximate cut posterior analyses.  The mixture approach allows repeated approximation of posterior distributions for different data based on a single mixture fit, which is important for model checks which aid in the decision of whether to ``cut".  
A semi-modular approach to likelihood-free inference
where feedback is partially cut is also developed.  The benefits of the method are illustrated in two challenging examples, 
a collective cell spreading model and a continuous time model for asset returns with jumps.


\end{abstract}

\section{Introduction}\label{sec:Intro}

Statisticians are increasingly using complex models which can be thought of as a collection of
coupled modules. The modules 
represent different aspects of our knowledge of the problem, and in a Bayesian analysis each module consists of likelihood
terms for different data sources and hierarchical prior terms for parameters or latent variables.  
There is much recent interest 
in ways to modify Bayesian inference so that it is fit for purpose when the model is misspecified, 
and for modularized Bayesian analyses so-called ``cutting feedback'' methods are common.  
The main goal of such methods is to ensure 
that inference about parameters of interest is not contaminated
by misspecified modules.  Cutting feedback approaches 
consider representations of the conventional Bayesian posterior distribution in terms of conditional or sequential
components, but then
modify certain terms for specification or computation of a modified posterior distribution.  
Existing cutting feedback methods have been developed in the context where the likelihood is tractable, 
and the purpose of this work is to develop suitable methods for intractable likelihood settings, 
where likelihood-free computational methods not requiring likelihood evaluations are used.  

To understand the motivation for cutting feedback methods, it is helpful to consider their
use in pharmacokinetic/pharmacodynamic (PK/PD)  modelling \citep{bennett+w01,lunn+bsgn09}.
This is one of the areas in which cutting feedback methods were first used and formalized.
In PK/PD applications, models with a two module structure are often considered.
One of the modules is a pharmacokinetic (PK) model
describing the evolution of a drug concentration in the blood stream, and the other is 
a pharmacodynamic (PD) model which describes
the effects of the drug on the body. There are module-specific data sources informing 
corresponding module parameters, and
the output of the PK module is used as an input to the PD 
module.   Often it can be difficult to specify the PD module adequately.  This can result in contamination of inferences
of interest due to the misspecification, as parameters of the PK module adapt to accommodate the misspecification
in the PD module.  
Cutting feedback methods have been used to prevent harmful effects arising from misspecification
in a situation like this, while still appropriately propagating uncertainty.  
There are many other applications of cutting feedback methods
-- see \cite{liu+bb09} and \cite{jacob+mhr17} for further discussion of these and modularized
Bayesian analyses more generally.  \cite{pompe+j21} and \cite{frazier+n22} 
have recently studied the theoretical behaviour of
the cut posterior distribution.

How are the goals of cutting feedback achieved?  
One way of formulating cutting feedback involves the modification of the steps of an MCMC sampling algorithm, 
where the cut posterior distribution is defined implicitly as the stationary distribution of the sampler.  
The ``cut'' function in the WinBUGS and OpenBUGS packages (see \cite{lunn+bsgn09} for details) is one way
to make this operational.
For a Gibbs sampling approach, the modified MCMC sampler draws parameter blocks from 
distributions obtained by removing some terms in the conventional posterior full conditional distributions.  
Removing misspecified terms when forming some of the full conditionals reduces the influence 
of misspecified components on the final inference.
The modified full conditional distributions are not in general consistent with any well-defined joint distribution
(see for example \cite{clarte+rrs20} and \cite{rodrigues+ns20} for a discussion of Gibbs sampling for
inconsistent conditionals in the likelihood-free inference setting).
When exact Gibbs sampling is intractable, it is natural to use Metropolis-within-Gibbs steps in detailed balance
with the modified conditional distributions.
In this case, the stationary distribution of the resulting Markov chain
may depend on the proposal distribution used \citep{woodard+cr13,plummer15}.     
To clarify the idea of cutting feedback, \cite{plummer15} and several other authors have considered
cutting feedback for a certain ``two module'' system discussed further in Section 2, which is general
enough to cover many applications of interest.
In this case, it is possible to characterize the cut posterior distribution explicitly.
This cut posterior distribution is not easy to sample from in most cases, 
due to the need to calculate or approximate a difficult normalizing constant.  

In this work we develop cutting feedback for likelihood-free inference.  
Likelihood-free inference methods are used with complex models where computation of the likelihood is impractical, but
where it is possible to simulate data under the model for any given value of the parameter.
We discuss these methods further in Section 3.
To address the computational challenges of cutting feedback in the likelihood-free setting, we use Gaussian
mixture approximations to the joint distribution of parameters and data summary statistics.  
We exploit the fact that marginal and conditional distributions of a Gaussian mixture are Gaussian mixtures 
to give explicit approximations to marginal or conditional posterior distributions for any summary statistic value
based on a single fitted mixture model,  
so that we can easily approximate cut posterior analyses and perform appropriate diagnostics.  
To the best of our knowledge,
our work is the first time that cutting feedback methods have been considered for likelihood-free inference.
A semi-modular inference approach where feedback
is partially cut is also developed, extending work by \cite{carmona+n20} and \cite{nicholls2022valid} 
to the case of likelihood-free inference.

In the next section we discuss cutting feedback for the two module system discussed in \cite{plummer15}.  This is followed
by a discussion of mixture approximations for likelihood-free inference.  Section 4 describes how mixture
approximations used in likelihood-free inference are able to address some of the computational difficulties of cutting feedback.  
Use of these mixture approximations for semi-modular inference methods is then discussed, and our methodology is applied
to two challenging examples in Section 6, a collective cell spreading model and a continuous time model for asset returns with jumps.  
Section 7 gives some concluding discussion.
  
\section{Cutting feedback}
  
\begin{figure}[h]
\centerline{\includegraphics[width=60mm]{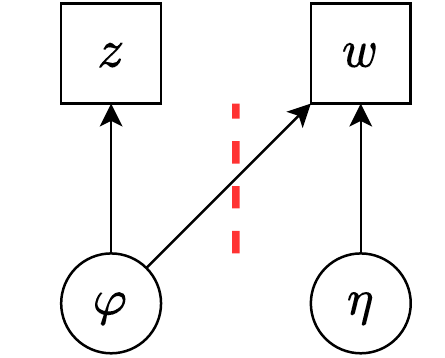}}
\caption{\label{two-module} Graphical representation of a two module system with cutting feedback.  The two modules are 
represented by the components on the left and right of the dashed line, which indicates
the cut.}
\end{figure}

To describe cutting feedback methods and clarify previous ``implicit'' definitions of cutting feedback in terms
of modified MCMC algorithms, \cite{plummer15} considered the two module system 
 represented graphically in Figure \ref{two-module}.  Our discussion in this paper will be restricted
 to the case of two modules, which is general enough to cover many applications of 
 interest.  For a discussion of cutting feedback in a more general 
 context, see \cite{lunn+bsgn09}.  
In Figure 1, the complete data, which we will write as $y$, 
is comprised of two data sources, $w$ and $z$.  The distribution of $z$ depends on parameter $\varphi$, and
the distribution of $w$ depends on parameters $\eta$ and $\varphi$.  We have two modules, the first consisting
of the prior for $\varphi$ and likelihood term for $z$ (module 1), and the second consisting of the prior for
$\eta$ and likelihood term for $w$, which also depends on $\varphi$ (module 2).  Suppose that we are concerned that module 2 is misspecified,
and that this might adversely affect inferences of interest.  The full posterior distribution can be written as
\begin{align}
  p(\varphi,\eta|y) & = p(\varphi|y)p(\eta|\varphi,w)  \nonumber \\
   & \propto p(\varphi|z)p(w|\varphi)p(\eta|\varphi,w)  \label{feedback}.
\end{align}
We have used the conditional independence of $\eta$ and $z$ given $w$ and $\varphi$ in the first line above, and 
the conditional independence of $w$ and $z$ given $\varphi$ in the second line.  In (\ref{feedback}) the term
$p(w|\varphi)$ is the ``feedback'' term which modifies the marginal posterior distribution for $\varphi$ from
$p(\varphi|z)$ to account for the information about $\varphi$ in the second module.  If the second module is
suspect, we might drop this term to obtain the so-called ``cut'' posterior distribution
\begin{align}
  p_{\text{cut}}(\varphi,\eta|y) & = p(\varphi|z)p(\eta|\varphi,y) \nonumber \\
   & = p(\varphi|z)p(\eta|\varphi,w)  \label{cutpost}
\end{align}
which rejects the feedback from module 2 about $\varphi$ while propagating uncertainty about $\varphi$ from
module 1 for making inferences about $\eta$. The red line in the figure indicates the cut.  

Computation for the cut posterior distribution is challenging.  We can write
\begin{align}
 p_{\text{cut}}(\varphi,\eta|y) & \propto p(\varphi)p(z|\varphi)\frac{p(\eta|\varphi)p(w|\eta,\varphi)}{p(w|\varphi)},
 \label{cutcomp}
\end{align}
but implementing an MCMC sampler is not easy from the likelihood and prior specification, because of the $p(w|\varphi)$
term on the right-hand side above which is usually not tractable. 
A number of computationally intensive methods for dealing with the intractable normalizing factor have
been suggested \citep{plummer15,jacob+oa17,liu+g20}.  \cite{yu+ns21} and \cite{carmona+n22} 
consider variational inference
methods which do not require approximation of the normalizing constant.  However, these previous works were concerned
with the case where the likelihood is tractable, and our main interest is in applications of cutting feedback to models
with intractable likelihoods, where likelihood-free methods are used.  

\section{Mixture approximations for likelihood-free inference}

The approach we use for likelihood-free inference will now be introduced, which allows us to easily approximate
cut posterior analyses. Write $y$ for the data and
$\theta$ for the parameters and we consider Bayesian inference with prior density $p(\theta)$.  If the likelihood
$p(y|\theta)$ is impractical to compute, then conventional Bayesian computation methods are inapplicable.  
However, there is now a large literature on likelihood-free inference methods able to perform Bayesian inference
using only model simulation, with approximate Bayesian computation (ABC) \citep{sisson+fb18} and synthetic likelihood 
\citep{wood10,price+dln16} being the traditional approaches.  More recently, approaches using flexible classification and regression methods from machine
learning \citep{gutmann+c16,raynal+mprre18,hermans+bl20,thomas+dckg21,pacchiardi+d22} are increasingly used.

\subsection{Mixture modelling of parameters and summaries}

Here we will use an approach to likelihood-free inference based on mixture modelling of the joint distribution of
the parameters and some summary statistics of the data.  This approach is developed 
in \cite{bonassi+yw11}, where they consider induced conditional distributions from a mixture model
as a form of nonlinear regression adjustment.  The method has been refined within a sequential Monte Carlo
framework in \cite{bonassi+w15}, although this extension is not helpful for the application considered
here where we require analytic forms for marginal and conditional posterior densities. 
Other authors have considered mixture models in likelihood-free inference as well, such as 
\cite{fan+ns13} who consider mixture of experts approximations to marginal summary statistics and copulas
to approximate the likelihood, and \cite{forbes+nna21} who consider mixture posterior approximations 
to define suitable functional discrepancy measures for ABC algorithms.  
\cite{papamakarios+m16}, \cite{lueckmann+gbonm17} and \cite{greenberg+nm19} have considered machine learning approaches based on mixture density networks.  
\cite{he+hy21} have recently developed
mixture variational posterior approximations for likelihood-free inference using a population Monte Carlo algorithm.

The simple mixture approach of \cite{bonassi+yw11} will be used here, since the ability to
approximate arbitrary marginal and conditional distributions from a single mixture fit 
given any subset of the summary
statistics, and for any values of those summary statistics, 
is crucial for the cutting feedback applications we describe.  The more
sophisticated mixture methods mentioned above either do not lead to closed form expressions for posterior
approximations or require expensive additional computations for each new posterior distribution
to be approximated, or both.  The method of \cite{bonassi+yw11} uses the following approach.
First, we suppose that we have some summary statistics of the data $S=S(y)$ available which are informative
about the model parameters.  The observed value of $S$ is written as $S_{\text{obs}}=S(y_{\text{obs}})$, 
where $y_{\text{obs}}$ is the observed value of $y$.  
We approximate $p(\theta|y_{\text{obs}})$ by $p(\theta|S_{\text{obs}})$.  If $S$ is a sufficient statistic then 
$p(\theta|y_{\text{obs}})$ and
$p(\theta|S_{\text{obs}})$ are the same, but a low-dimensional sufficient statistic 
is rarely available in likelihood-free inference applications.  
It is important that $S$ is low-dimensional if we are to estimate the distribution of $S$ 
from simulated data.  Having chosen $S$, 
we simulate samples $(\theta_i,S_i)$, $i=1,\dots, n$ from $p(\theta)p(S|\theta)$.  Next, we fit a Gaussian
mixture model to the simulated data.  Writing $U=(\theta,S)$, this gives an estimate $\widetilde{p}(u)=\widetilde{p}(\theta,S)$
for the joint density.  The 
posterior density $p(\theta|S_{\text{obs}})$ is just the conditional density of $\theta$ given $S=S_{\text{obs}}$ in
$p(\theta,S)$, which we can approximate by the corresponding conditional density in the mixture model,
$\widetilde{p}(\theta|S=S_{\text{obs}})$.  

An observation which is crucial later is that any marginal distribution of $\widetilde{p}(u)$ is a Gaussian
mixture model, and any conditional distribution is also a Gaussian mixture.  So from the mixture approximation
$\widetilde{p}(u)$, we can obtain a closed form approximation to any marginal or conditional posterior distribution conditional on any subset
of summary statistics, and we can do this for any value of the summary statistics based on a single
mixture model fit.  To make this explicit, suppose our mixture approximation is 
\begin{align*}
  \widetilde{p}(u) & = \sum_{j=1}^J w_{j,u}\phi_{j,u}(u),
\end{align*}
where $J$ is the number of mixture components, $w_{j,u}$, $j=1,\dots, J$ are non-negative mixing weights summing
to $1$, and $\phi_{j,u}(u)$ is a multivariate normal component density with mean $\mu_{j,u}$ and covariance matrix
$\Sigma_{j,u}$, $j=1,\dots, J$.  Consider a subvector $V=(X,W)$ of $U$ where $X$ and $W$ are disjoint.  
In our later applications to cutting feedback, where we approximate marginal and conditional posterior
distributions, $X$ will be a subset of the model parameters, and $W$ can consist of both parameters and summary
statistics.  
Write the marginal distribution of $V$ for the mixture component $\phi_{j,u}(u)$ as 
$\phi_{j,v}(v)$, which is multivariate normal with mean vector $\mu_{j,v}$ and covariance matrix $\Sigma_{j,v}$, $j=1,\dots, J$.  
Write $\mu_{j,v}$ in partitioned form as $\mu_{j,v}=(\mu_{j,x}^\top,\mu_{j,w}^\top)^\top$, and
\begin{align*}
  \Sigma_{j,v} & = \left[\begin{array}{cc} 
    \Sigma_{j,x} & \Sigma_{j,xw} \\
    \Sigma_{j,xw}^\top & \Sigma_{j,w} \end{array}\right],
\end{align*}
where the partitioning is conforming to the partition of $V$ as $(X,W)$.  
We also write the marginal density of $W$ for component $j$ of the mixture as
$\phi_{j,w}(w)$, $j=1,\dots, J$.  
The conditional density of $X|W$ is a Gaussian
mixture, 
\begin{align*}
  \widetilde{p}(x|w) = \sum_{j=1}^J  w_{j,x|w} \phi_{j,x|w}(x),
\end{align*}
where
$$w_{j,x|w} = \frac{w_{j,u}\phi_{j,w}(w)}{\sum_{l=1}^J w_{l,u}\phi_{l,w}(w)},$$
and $\phi_{j,x|w}(x)$ is multivariate normal with mean and covariance matrix
$$\mu_{j,x|w}=\mu_{j,w}+\Sigma_{j,xw}\Sigma_{j,w}^{-1}(w-\mu_{j,w}),$$
and
$$\Sigma_{j,x|w}=\Sigma_{j,x}-\Sigma_{j,xw}\Sigma_{j,w}^{-1}\Sigma_{j,xw}^\top,$$
respectively.

\subsection{Mixture approximations and cutting feedback}

Now we consider the issue of cutting feedback for likelihood-free inference.  
Suppose the summary statistics are partitioned as $S=(S_1^\top, S_2^\top)^\top$, and 
write the corresponding partition of $S_{\text{obs}}$ as $S_{\text{obs}}=(S_{\text{obs},1}^\top,S_{\text{obs},2}^\top)^\top$.  
We wish to base inference about parameters $\varphi$ only on $S_1$, because we are worried
that the summaries $S_2$ adversely affect inference about $\varphi$.  
The information in $S_2$, however,
may be valuable for inference about the remaining parameters $\eta$.  
Example 5 of \cite{sisson+fb18intro} illustrates a simple situation where model misspecification 
can lead to summary statistics with conflicting information.
In most likelihood-free inference applications 
there is no graphical structure to the model such as in Figure \ref{two-module}, 
and if we regard $S_1$ and $S_2$ as data sources associated with two modules, there
is no conditional independence between them given the parameters.  However, similar
to (\ref{cutpost}) it is still useful to define a cut posterior
which ignores $S_2$ in inference about $\varphi$.  For any value of $S$, we write 
\begin{align}
 p_{\text{cut}}(\theta|S) & = p(\varphi|S_1)p(\eta|\varphi,S).  \label{cut-lfi}
\end{align}

As discussed in the last section, given a mixture approximation $\widetilde{p}(\theta,S)$ to $p(\theta,S)$, 
an analytic form for the conditional densities for $\varphi|S_1$ and $\eta|\varphi,S$ can be written down.   
Then  (\ref{cut-lfi}) 
can be approximated by 
\begin{align}
 \widetilde{p}_{\text{cut}}(\theta|S) & = \widetilde{p}(\varphi|S_1) \widetilde{p}(\eta|\varphi,S),  \label{cutapprox}
\end{align}
where $\widetilde{p}(\varphi|S_1)$ and $\widetilde{p}(\eta|\varphi,S)$ are the conditional densities induced
from $\widetilde{p}(\theta,S)$.  Monte Carlo summarization of the cut posterior approximation (\ref{cutapprox}) is 
easy, since we just need to do sequential simulation from two Gaussian mixture models.
In the next subsection we discuss methods for deciding whether or not to cut, where it is required to
compute certain posterior distributions repeatedly for different data simulated under a reference distribution; 
the mixture approach can perform the required computations based on only a single mixture model fit.  


An interesting case of the framework above is when $S=y$, and the likelihood is tractable but we wish to base inference
about $\varphi$ only on a low-dimensional summary statistic $S_1$ for which $p(S_1|\varphi)$ is not analytically
available.  The approximate Bayesian forecasting approach considered in 
\cite{frazier+mmm19} falls into this framework, where the authors consider inferring the parameter $\varphi$ in a state space model
using ABC with summary statistics, and then for ABC draws for $\varphi$ they sample the conditional posterior
distribution of latent states $\eta$ given $y$ using a particle filter in order to produce 
forecasts.  Only filtering, and not smoothing,
is needed for predictive inference.  Here there is no need for ABC approximations in inferring the conditional
posterior distribution of the states $p(\eta|\varphi,y)$, but ABC methods can be useful for inferring the parameters $\varphi$.  
\cite{frazier+mmm19} use predictive criteria for the choice of summary statistics, which might be particularly beneficial
in the case of misspecification.  

\subsection{Deciding whether or not to cut}

When considering the use of cut methods we may need to decide whether to use the cut or conventional posterior.   
It is easy to see that the Kullback-Leibler divergence between the cut posterior distribution
(\ref{cut-lfi}) and the full posterior distribution is the Kullback-Leibler divergence between their marginal posterior distributions
for $\varphi$:  
\begin{align*}
  \text{KL}(p(\theta|S)\|p_{\text{cut}}(\theta|S)) & = \text{KL}(p(\varphi|S)\|p(\varphi|S_1)).
\end{align*}
This follows from the fact that the conditional posterior distribution for $\eta$ given $\varphi$ is the
same in both distributions.  For a proof of this see Lemma 1 of \cite{yu+ns21}.  
Let us write
\begin{align}
 G(S_2|S_1) & = \text{KL}(p(\varphi|S)\|p(\varphi|S_1))  \label{prior-post}
\end{align}
The statistic $G(S_2|S_1)$ can be thought of as a prior-to-posterior divergence, in the situation where $S_1$ is known
when forming the prior but before we know $S_2$.  
\cite{nott+wee20} consider prior-data conflict checks based on such prior-to-posterior divergences, and
\cite{yu+ns21} consider the use of these checks for deciding whether or not to cut feedback in a 
Bayesian analysis with tractable likelihood.
Similar conflict checks were developed in likelihood-free inference in \cite{chakraborty+ne21}, 
although \cite{chakraborty+ne21} do not consider cutting feedback methods.  

We can approximate (\ref{prior-post}) by replacing $p(\varphi|S)$ and $p(\varphi|S_1)$
by $\widetilde{p}(\varphi|S)$ and $\widetilde{p}(\varphi|S_1)$ respectively.  Since both of these densities are 
Gaussian mixtures, we can also make use of closed form approximations to Kullback-Leibler divergences between
mixtures \citep[Section 7]{hershey+o07} to obtain an approximation $\widetilde{G}(S_2|S_1)$
to $G(S_2|S_1)$.  For more details, see \cite{chakraborty+ne21}.
To decide whether or not to cut, we compare the statistic $\widetilde{G}(S_{\text{obs},2}|S_{\text{obs},1})$ to
the distribution of $\widetilde{G}(S_2'|S_{\text{obs},1})$, where $S_2'$ is a draw from $p(S_2|S_{\text{obs},1})$, the conditional prior predictive
for $S_2$ given $S_1=S_{\text{obs},1}$.  
Simulation from $p(S_2|S_{\text{obs},1})$ can be approximated by simulation from $\widetilde{p}(S_2|S_{\text{obs},1})$
if necessary. 
If $\widetilde{G}(S_{\text{obs},2}|S_{\text{obs},1})$ lies out in the tails of this distribution, it says that
the cut posterior distribution has changed an unusually large amount from the full posterior distribution if the model
is correct for the observed $S_2$.  Precisely, we consider the tail probability
\begin{align}
  p & = P(\widetilde{G}(S_2'|S_{\text{obs},1})\geq \widetilde{G}(S_{\text{obs},2}|S_{\text{obs},1})), \label{tail}
\end{align}
where $S_2'\sim p(S_2|S_{\text{obs},1})$. 
If this tail 
probability is small, the change from the cut posterior distribution to the full posterior
distribution is unusually large for the observed data compared to what is expected if the model is correct. 
In approximating (\ref{tail}) by Monte Carlo simulation, computation of the approximate Kullback-Leibler divergence
between the full and cut marginal posterior distributions for $\varphi$ can be done for repeated simulated
summary statistics $S_2'$ under the reference distribution based on the same single mixture fit that was used
for the cut model computations for the observed summary statistics.

\section{Semi-modular inference}

As a generalization of cut posterior approaches, \cite{carmona+n20} introduced semi-modular inference, which
gives a mechanism for partially cutting feedback.  They consider the two module system of Figure 1, and
suggest using some of the full module structure in making inference
about $\varphi$.  An influence parameter $\gamma\in [0,1]$ tempers the influence of the possibly misspecified module on inference about $\varphi$, whereas the conditional posterior distribution for $\eta$ given $\varphi$ is that of the full posterior.  

The construction of \cite{carmona+n20} considers a two stage approach.  First, a ``power posterior distribution'' is constructed
for inference about $\varphi$ and an auxiliary replicate parameter $\widetilde{\eta}$ of $\eta$, given the data
$z$ and $w$.  The use of similar power posterior distributions \citep{bissiri+hw16,grunwald+v17,miller+d19} 
for robust Bayesian inference originates
outside the modular inference context.  
Following \cite{carmona+n20}, we use the notation $p_{\text{pow},\gamma}(\varphi,\widetilde{\eta}|y)$ for the power posterior for $(\varphi,\widetilde{\eta})$  
with influence parameter $\gamma$, and this is defined to be
\begin{align*}
  p_{\text{pow},\gamma}(\varphi,\widetilde{\eta}|y) & \propto p(z|\varphi)p(w|\varphi,\widetilde{\eta})^\gamma p(\varphi,\widetilde{\eta}).
\end{align*}
The influence parameter tempers the likelihood term from the second module, reducing its influence.  Next, the semi-modular
posterior distribution is defined as 
\begin{align*}
  p_{\text{smi},\gamma}(\varphi,\eta,\widetilde{\eta}|z,w) & = p_{\text{pow},\gamma}(\varphi,\widetilde{\eta}|z,w)p(\eta|w,\varphi),
\end{align*}
and inference about $\theta=(\varphi,\eta)$ is achieved by integrating out $\widetilde{\eta}$.  
Setting $\gamma=0$, the cut posterior for $\theta$ is obtained, and setting $\gamma=1$ gives the full posterior.  
Hence the semi-modular approach interpolates between the cut and full posterior distributions.  \cite{carmona+n20} 
suggest choosing $\gamma$ using predictive methods.  More recently, \cite{nicholls2022valid} consider validity
of semi-modular inference in a generalized Bayesian inference framework, and consider alternative forms 
of semi-modular inference.  We consider one more alternative below for the likelihood-free setting.

\subsection{Likelihood-free semi-modular inference}

We now develop an alternative semi-modular posterior 
construction for the likelihood-free setting, using a method for constructing marginal
inferences for $\varphi$ inspired by linear opinion pooling \citep{stone61}.  
Linear opinion pooling \citep{stone61} combines
distributions representing opinions of different experts using a mixture model.  We define a 
semi-modular marginal posterior distribution for $\varphi$ as a mixture between the cut and full marginal
posterior distributions with mixing weight $\gamma$.  Such an approach is natural 
when mixture approximations are used for computation.  
In that case, the cut marginal posterior
approximation for $\varphi$ and the full marginal posterior approximation for $\varphi$ are Gaussian mixtures, and a
mixture of them is also a Gaussian mixture.  This makes it easy to approximate our 
proposed semi-modular posterior distribution
with an explicit form, and we suggest a convenient way to choose the influence parameter $\gamma$ based on
calculations similar to those used for the prior-data conflict checks in Section 3.3.  

Define 
\begin{align}
  p_{\gamma}(\varphi|S) & = \gamma p(\varphi|S)+(1-\gamma)p(\varphi|S_1),  \label{smivarphi}
\end{align}
where $\gamma\in [0,1]$ is the influence parameter.  We define a semi-modular posterior distribution for 
$\theta=(\varphi,\eta)$ by
\begin{align}
  p_{\text{smi},\gamma}(\theta|S) & = p_{\gamma}(\varphi|S)p(\eta|\varphi,S). \label{smi}
\end{align}
Using the mixture posterior approximations of Section 3 for computation, the approximations to the posterior
densities of $\varphi$ given $S_1$ and $\varphi$ given $S$ respectively are written
\begin{align}
 \widetilde{p}(\varphi|S_1) & = \sum_{j=1}^J w_{j,\varphi|S_1}\phi_{j,\varphi|S_1}(\varphi), \label{varphiS1}
\end{align}
and
\begin{align}
  \widetilde{p}(\varphi|S) & = \sum_{j=1}^J w_{j,\varphi|S}\phi_{j,\varphi|S}(\varphi). \label{varphiS}
\end{align}
Similarly, write the mixture approximation to the density of $\eta|\varphi,S$ as 
\begin{align*}
  \widetilde{p}(\eta|\varphi,S) & =\sum_{j=1}^J w_{j,\eta|\varphi,S}\phi_{j,\eta|\varphi,S}(\varphi).
\end{align*}

We approximate (\ref{smivarphi}) by 
\begin{align}
  \widetilde{p}_\gamma(\varphi|S) & = \gamma \widetilde{p}(\varphi|S)+(1-\gamma) \widetilde{p}(\varphi|S_1), \label{smivarphimix}
\end{align}
and the semi-modular posterior (\ref{smi}) by
\begin{align}
  \widetilde{p}_{\text{smi},\gamma}(\varphi,\eta|S)=\widetilde{p}_\gamma(\varphi|S)\widetilde{p}(\eta|S).  \label{smimix}
\end{align}
$\widetilde{p}_\gamma(\varphi|S)$ is itself a Gaussian mixture, with $2J$ components, and component densities
$$\{\phi_{j,\varphi|S_1}(\varphi),\, \phi_{j,\varphi|S}(\varphi): j=1,\dots, J\}$$ 
and corresponding mixing weights 
$$\{\gamma w_{j,\varphi|S},\, (1-\gamma)w_{j,\varphi|S_1}:  j=1,\dots, J\}.$$
Hence both terms on the right-hand side of (\ref{smimix}) are Gaussian mixtures.  Similar to the cutting feedback
case of Section 3.2, Monte Carlo summarization of the semi-modular approximation is easy, involving sequential simulation
from two mixture models.

\subsection{Choosing the influence parameter}

We now outline a convenient approach to the choice of the influence parameter $\gamma$.  We write 
\begin{align*}
  G_{\gamma}(S) & = \text{KL}(p_{\gamma}(\varphi|S)\|p(\varphi|S_1))
\end{align*}
for the Kullback-Leibler divergence between the semi-modular marginal posterior distribution for $\varphi$ and $p(\varphi|S_1)$.  
If $\gamma=1$, then this is the conflict checking statistic $G(S_2|S_1)$ considered in Section 3.3.  
Similar to the discussion of Section 3.3, we can replace $p_{\gamma}(\varphi|S)$ and $p(\varphi|S_1)$ by their
mixture approximations $\widetilde{p}_\gamma(\varphi|S)$ and $\widetilde{p}(\varphi|S_1)$, and use a closed-form
approximation to Kullback-Leibler divergences between mixtures \citep[Section 7]{hershey+o07} to obtain
an approximate statistic $\widetilde{G}_\gamma(S)$ that is easy to compute.  

Define a tail probability 
\begin{align}
  p(\gamma)=P(\widetilde{G}(S_2'|S_{\text{obs},1})\geq \widetilde{G}_\gamma(S_{\text{obs}})), \label{pgamma}
\end{align}
where $S_2'\sim p(S_2|S_{\text{obs},1})$.  Simulation
of $S_2'$ can be approximated by simulation from $\widetilde{p}(S_2|S_{\text{obs},1})$ if needed.  
We propose to choose $\gamma$ as the largest value $\gamma'$ such that $p(\gamma')>\alpha$, where 
$\alpha$ is a cutoff for a measure of surprise such as $0.05$.  Finding $\gamma'$ can be done by computing $p(\gamma)$ for
$\gamma$ on a grid.  The intuitive meaning of choosing $\gamma$ in this way is the following.  If there is no conflict 
at level $\alpha$
according to the check of Section 3.3, 
then we choose $\gamma=1$ and we use the full posterior.  If there is a conflict, then we back off from $\gamma=1$ 
to a smaller value such that the conflict would be avoided if $\widetilde{p}_\gamma(\varphi|S)$ 
had been the full posterior marginal for $\varphi$.  The idea is to use as much of the full posterior information as possible, 
subject to retaining an interpretation for the inference that is not in conflict with that based only on the
summary $S_1$.

\section{Examples}

We consider two examples.  The first concerns a collective cell spreading model.  
A common use for cutting feedback methods is to explore whether misspecification of one module
impacts inference about certain parameters.  If cut and full posterior inferences have a similar
interpretation, this might be reassuring that inferences of interest are not sensitive to the misspecification.  
In the cell spreading model it has been noted in past work by \cite{frazier+d21} 
that the cell interaction component 
fails to capture some aspects of the observed data, and for this example we use cutting feedback 
to demonstrate that this inadequacy does not affect inference about cell proliferation.  
The example also illustrates the usefulness of the mixture modelling approach for exploring the informativeness
of summary statistics for inference about different parameters in a computationally thrifty way.
Our second example considers time series models for asset returns with jumps.  We 
start by considering a continuous time model and 
explore cutting feedback so that jump parameters are estimated 
using only summary statistics on high frequency intra-day returns.  
This results in different inferences about the jump parameters compared to
the ordinary posterior where summary statistics incorporate both information from both daily
and intra-day returns.  A similar discrete-time model is then discussed,
where the performance of full, cut and semi-modular posterior distributions are explored for forecasting purposes, and
where cut and semi-modular approaches improve on the conventional posterior distribution for predictive
purposes.

\subsection{Collective cell spreading}

Our first example considers a model developed in \cite{browning18} for collective cell-spreading.  Their model is useful in
applications to understanding skin cancer growth and wound healing.  Misspecification for this model was discussed in
the supplementary material of \cite{frazier+d21}, and we apply cutting feedback methods to understand the
effect of this misspecification on inference.  The model has three unknown parameters, $\theta = (m, \rho, \gamma_b)^\top$.  The rates of motility (cell movement) and proliferation (cell birth) are given by $m$ and $\rho$, respectively.  The parameter $\gamma_b$ is part of a Gaussian kernel used to measure the closeness of cells.  The prior distribution for $\theta$ has independent 
uniform components, $\rho \sim U(0,10)$, $m \sim U(0,0.1)$ and $\gamma_b \sim U(0,20)$. The reader is referred to \cite{browning18} for a detailed discussion of the model.

The summary statistics are as follows:
\begin{enumerate}
	\item The number of cells at 12, 24 and 36 hours ($S_1$).
	\item The pair correlation computed at 12, 24 and 36 hrs ($S_2$).
\end{enumerate}
The summary statistic vector $S_1$ is intended to be informative about $\rho$, and $S_2$ is intended to be informative about $m$ and $\gamma_b$.  It is reasonable to suspect that the observed number of cells can be recovered for appropriate values of the model parameter.  However, it is more challenging to capture the spatial dependence of the cell population, i.e.\ how the cells interact with each other.  Accurate estimation of $\rho$ is important as cell proliferation drives cancer growth, and cancer treatments would aim to reduce this parameter.  Therefore, there is an interest in inferring $\rho$ in a way that is robust to potential misspecification of the cell interaction component of the model.  We consider a simulated dataset first where the model is correctly specified, using 
$\theta = (1, 0.04, 6)$.  For the real data, \cite{frazier+d21} show that the model does not capture the way that 
the observed pair correlation changes over time.

\begin{figure}[h]
 	\centering
 	\includegraphics[scale=0.75]{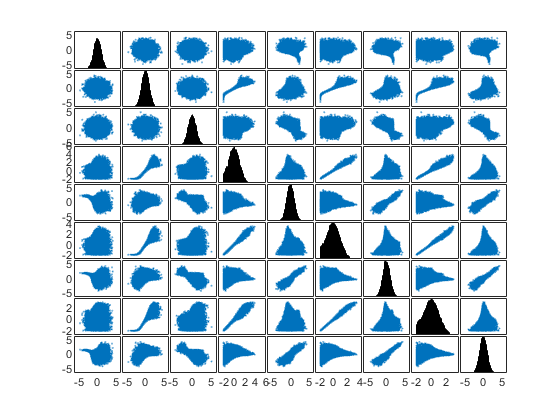}
 	\caption{Marginal distributions and bivariate scatterplots based on the joint distribution of parameters and summary statistics generated from prior predictive distribution (after a marginal transformation of the parameters and summary statistics).}
 	\label{fig:scatter}
 \end{figure}

We generate $N = 10^5$ simulations from the prior predictive distribution.  In an effort to improve the Gaussian mixture model (GMM) fit, we first transform each marginal summary statistic and parameter distribution using the probability integral transform to uniform, and then push the transformed samples through the standard normal quantile function.  For the prior distribution of the parameters, the distribution function has an analytic form.  For the summary statistics, the distribution function is estimated by kernel density estimation, and so the marginal distribution of the transformed summaries is only approximately standard normal.  Based on Figure \ref{fig:scatter}, even after transforming the marginal distributions, the dependence structure is complex, both between parameters and summary statistics, and between the summary statistics themselves.  Hence this represents a challenging application for the GMM approach.

A GMM with 10 components was fitted to the transformed prior predictive samples.  In the GMM we assumed 
unrestricted component covariance matrices, and estimation was performed using an EM algorithm.
The fitted GMM 
was then used to approximately sample the posterior density $p(\theta|S_1,S_2)$ and the cut posterior density $p_{\text{cut}}(\theta|S_1,S_2) = p(\rho|S_1)p(m,\gamma_b|\rho,S_1,S_2)$.  The  parameter samples generated from the fitted GMM were passed through the inverse transform to generate samples from the approximate posterior distributions on the original parameter space.  

\begin{figure}[h]
	\centering
	\includegraphics[width=160mm,height=60mm]{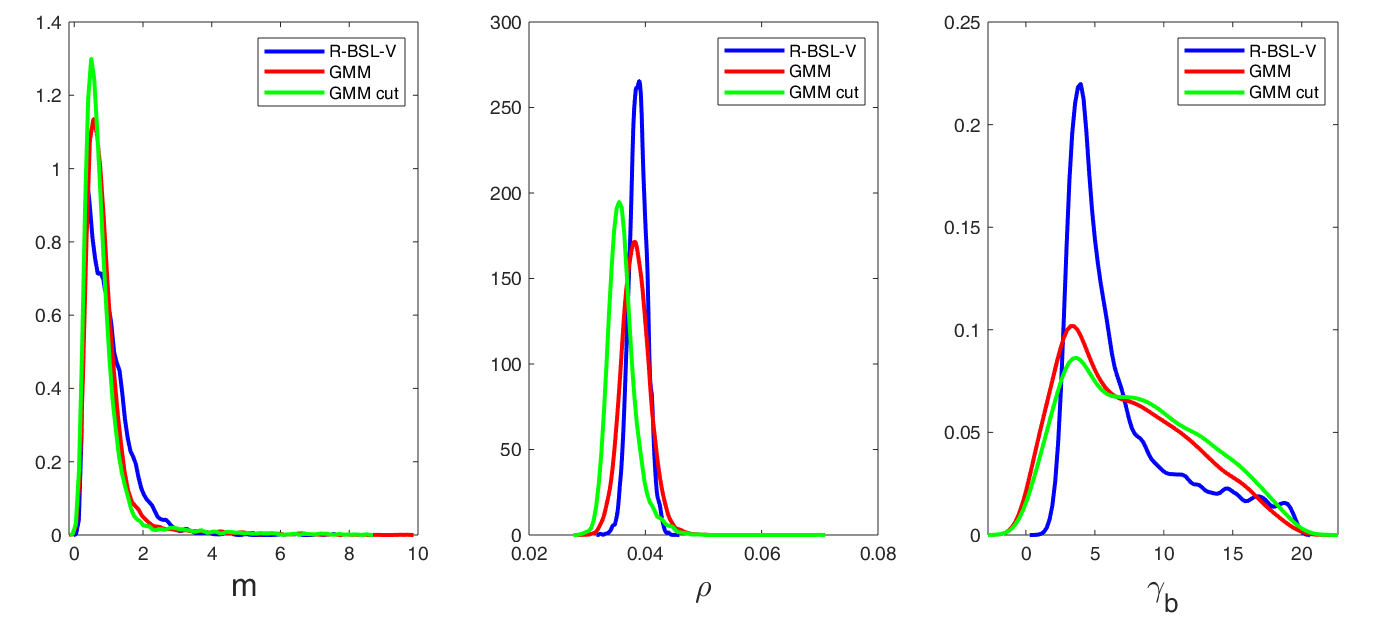}
	\caption{Estimates of marginal posterior densities for $m$, $\rho$ and $\gamma_b$ obtained using
	robust Bayesian synthetic likelihood with variance inflation (R-BSL-V), GMM approximation to the full posterior (GMM) and GMM approximation to the cut posterior (GMM cut) for the simulated data for collective cell spreading model.}
	\label{fig:cell_results_sim}
\end{figure}

\begin{figure}[h]
	\centering
	\includegraphics[width=160mm,height=60mm]{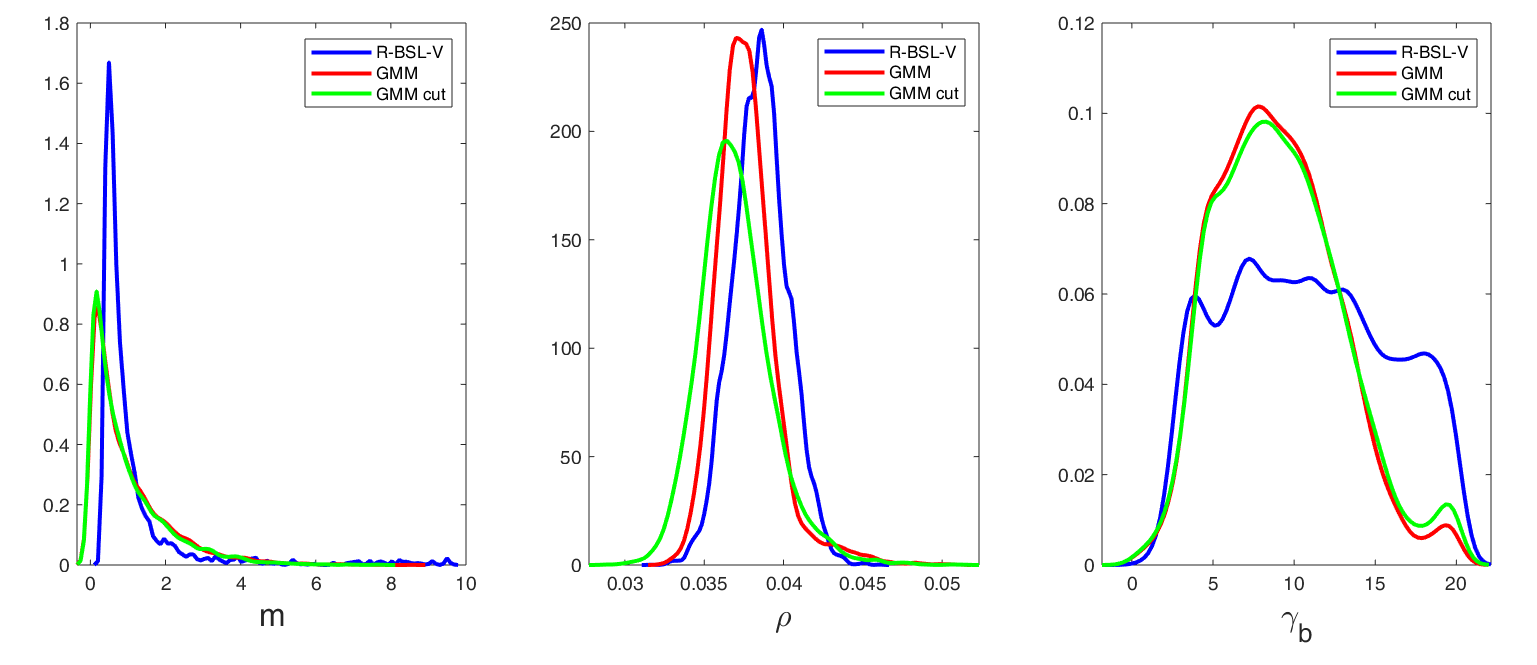}
	\caption{Estimates of marginal posterior densities for $m$, $\rho$ and $\gamma_b$ obtained using
	robust Bayesian synthetic likelihood with variance inflation (R-BSL-V), GMM approximation to the full posterior (GMM) and GMM approximation to the cut posterior (GMM cut) for the real data for collective cell spreading model.}
	\label{fig:cell_results_real}
\end{figure}

The marginal posterior estimates for the simulated and real datasets are shown in Figures \ref{fig:cell_results_sim} and \ref{fig:cell_results_real}, respectively.  We also include the estimates from \cite{frazier+d21} using their robust BSL method with variance inflation, which should be robust to potential misspecification of the spatial dependence of cells, in the sense that it can reduce the influence of summaries that the model is not compatible with.  It can be seen that the cutting feedback GMM posterior approximation does not differ greatly from the GMM posterior approximation, even for the real data.  This is consistent with the results obtained in \cite{frazier+d21}.  This indicates that the inability to recover $S_2$ is not adversely affecting inference about $\rho$. The GMM posterior approximations are surprisingly accurate given the greatly reduced number of model simulations used compared to BSL.

\begin{figure}[h]
	\centering
	\includegraphics[width=160mm,height=60mm]{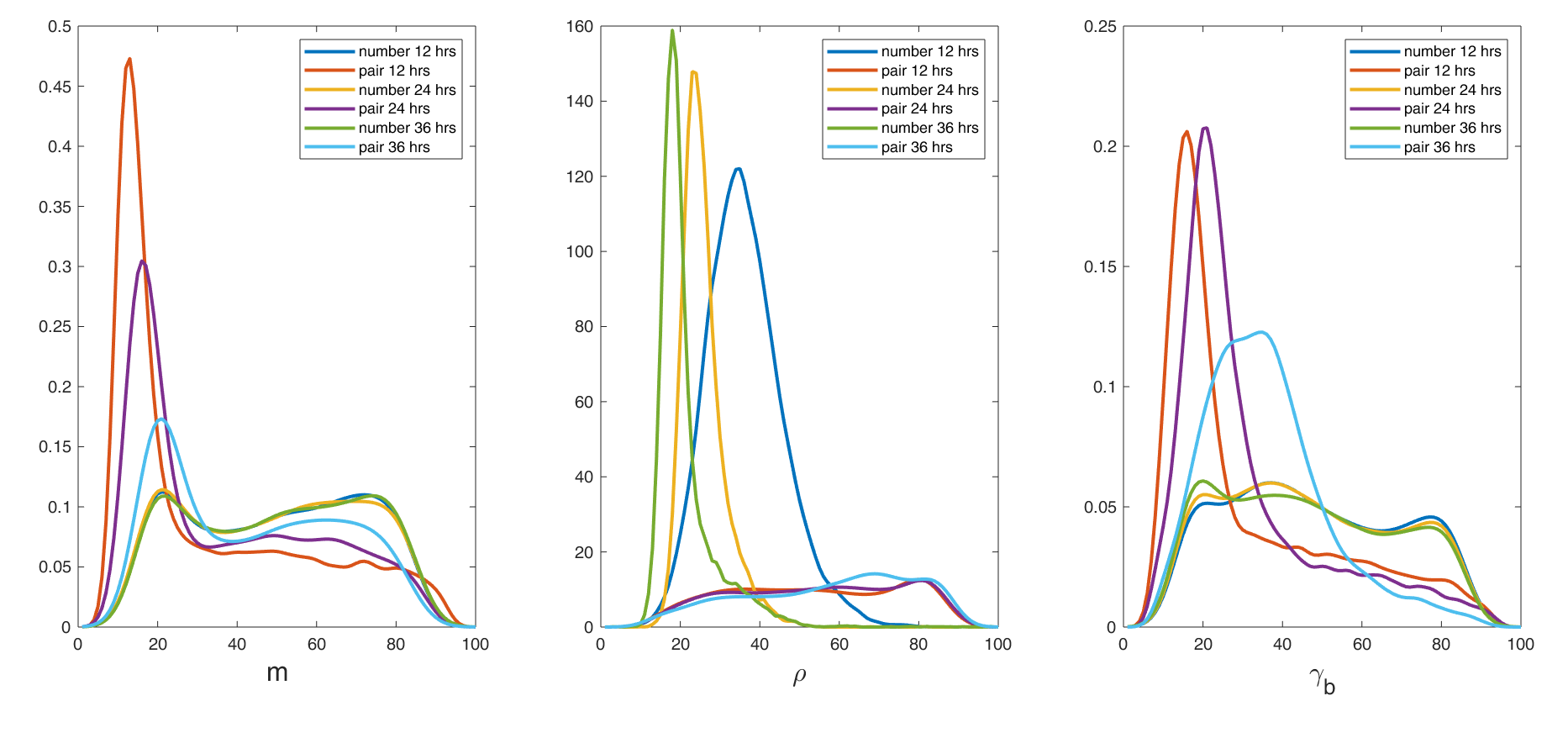}
	\caption{Sensitivity of posterior approximations to different summary statistics based on simulated data for the collective cell spreading example.  Estimated marginal posterior densities are shown for $m$, $\rho$ and $\gamma_b$.  The different lines in each plot are estimates obtained by conditioning on different scalar summary statistics.  }
	\label{fig:cell_results_sensitvity}
\end{figure}

Another potential advantage of the GMM approach is that we can thriftily explore the sensitivity of different summary statistic choices on each parameter.  Figure \ref{fig:cell_results_sensitvity} illustrates this using the simulated data, where marginal posterior densities for different parameters are estimated using each summary statistic individually.  
It can be seen that $\rho$ is informed by the number of cells at each time point, and the pair correlation statistics provide no information.  From this perspective, it is not surprising that $\rho$ is not adversely influenced by potential incompatibility of $S_2$, since the parameter is not sensitive to these statistics.  The estimated posterior of $\rho$ conditional on only the number of cells at 36 hrs is similar to the posterior approximation conditional on all the statistics.  The results suggest that it is only necessary to include the number of cells at the final time point in the summary statistic vector.  It is evident that $m$ and $\gamma_b$ are informed by $S_2$ and not $S_1$.  In the case of $m$, there does seem to be a benefit in including all time points, as the posterior approximation with the three pair correlation statistics is more concentrated than with any individual time point. 

\subsection{Continuous time model for asset returns with jumps}

Our next example considers a continuous time model for asset returns with jumps and 
 illustrates the use of our conflict checking approach for deciding whether or not to cut feedback, 
 as well as our proposed semi-modular method.  Here the use of a cut posterior distribution results
 in different inferences to those of the full posterior distribution.
Let $P_t$ denote the instantaneous price of an asset at time $t\ge0$.  
For $p_t=\ln P_t$, suppose that $p_t$ evolves according to the bivariate jump-diffusion process
\begin{flalign}
\dt p_{t} &=\mu_p \dt t+\exp \left(V_{t} / 2\right) \dt W_{t}^{p}+\dt J_{t}^{p} \label{returns} \\ 
\dt V_{t} &=\kappa\left(\alpha-V_{t}\right) \dt t+\sigma_{v} \dt W_{t}^{v} \label{volatility} \\ 
\dt J_{t}^{p} &=Z_{t} \dt N_{t}, \quad Z_{t} \sim N\left(\mu_{z}, \sigma_{z}^{2}\right), \label{jump}
\end{flalign}
where $W_t^j$, $j\in\{v,p\}$, are correlated Brownian motion processes, with instantaneous correlation $\rho$, $N_t$ is a counting process, $V_t$ is a latent volatility process $J_t^p$ is a process of unobservable jumps, and
$\mu_p$, $\kappa$, $\alpha$, $\sigma_v$, $\mu_z$ and $\sigma_z$ are unknown parameters.    

The above model is similar to the model of \cite{creel+k15}, however, we model the jump process via a conditionally deterministic Hawkes process \citep{ait-sahalia+cl15,maneesoonthorn+fm17} for which
\begin{align}
\text{Pr}\left(\dt N_{t}=1\right) & =\delta_{t} \dt t+o(1), \;\;\dt \delta_t =  (d+\beta \delta_t) \dt t +\tau \dt N_{t}, 
\label{hawkes} 
\end{align}
where $d$, $\beta$ and $\tau$ are positive unknown parameters.  
We refer to (\ref{returns}) as the returns model, (\ref{volatility}) as the volatility model, and (\ref{jump}) and (\ref{hawkes}) 
as the jump
model.  The unknown parameters are collected as $\theta=(\varphi^\top,\eta^\top)^\top$, where 
$$\varphi=(\mu_z,\sigma_z,d,\beta,\tau)^\top,\quad \eta=(\mu_p,\kappa,\alpha,\sigma_v,\rho)^\top.$$  The parameters $\varphi$ control the jump dynamics, while the parameters $\eta$ 
are those appearing in the returns and volatility models.

For a general value of $\theta$, and a given sequence of observed log-returns $\{r_t=p_t-p_{t-1}:t\ge1\}$, the likelihood associated with the model in \eqref{returns}-\eqref{hawkes} is intractable. This intractability is due to the presence of the unobservable state variables $V_t$, the latent volatilities, and $J_t$, the unobservable jumps. These variables must be integrated out of the measurement equation for the observables to obtain a likelihood that depends only on the observable data. This integration is made even more difficult by the fact that the transition equations for the latent states do not admit closed-form densities, due to their continuous-time evolution, and must generally be approximated. In contrast, simulation-based methods bypass calculation of the likelihood function by simulating data directly from the model.

We consider the case where
the researcher is uncertain of the error specification in the volatility/returns equations, 
and/or the specification of the jump dynamics.
Let us focus on the jump dynamics as a single module, and the volatility/returns specification as a separate module. 
If intra-day returns are available, inference on the parameters $\varphi$ governing the jump dynamics can proceed by 
``cutting'' the link with the returns and volatility equations.  
We observe daily log returns $r_t=p_t-p_{t-1}$ at integer times 
$t=1, \dots ,T$, and, for each $t$, we observe $M$ equally spaced intra-day returns $r_{t,i}$, with $i=1,\dots,M$.  Define the bipower variation, $\mathrm{BV}_t$, and jump-variation, $\mathrm{JV}_t$, as 
$$
\mathrm{B V}_{t}:=\frac{\pi}{2}\left(\frac{M}{M-1}\right) \sum_{i=2}^{M}\left|r_{t, i} r_{t,(i-1)}\right|,\quad \mathrm{JV}_{t}:=\max \left\{\mathrm{RV}_{t}-\mathrm{BV}_{t}, 0\right\}, 
$$  where  $\mathrm{RV}_t$ denotes realized volatility $\mathrm{RV}_{t}=\sum_{i=1}^{M} r_{t, i}^{2}$. ABC inference on the jump-dynamics can be carried out using the following summary statistics (Frazier, et al., 2019):
$$
S_{1,1}:=\frac{1}{T} \sum_{t=1}^{T} \operatorname{sgn}\left(r_{t}\right) \sqrt{\mathrm{JV}_{t}}, \quad S_{1,2}:=\frac{1}{T} \sum_{t=1}^{T}\left(\mathrm{JV}_{t}-\overline{\mathrm{JV}}_{t}\right)^{2}, $$
$$S_{1,3}:=\frac{1}{T} \sum_{t=2}^{T}\left(\mathrm{JV}_{t}-\overline{\mathrm{JV}}_{t}\right)\left(\mathrm{JV}_{t-1}-\overline{\mathrm{JV}}_{t}\right),
$$where $\overline{\mathrm{JV}}_{t}=T^{-1}\sum_{t=1}^{T}\mathrm{JV}_t$. 
We also define $S_{1,4}$ and $S_{1,5}$ as the sample skewness and kurtosis respectively of $\log \mathrm{BV}_t$. 
The summary statistic vector $S_1$ used for inference on $\varphi$ in the ``cut'' posterior is $S_1=(S_{1,1},S_{1,2},S_{1,3},S_{1,4},S_{1,5})^\top$.

The unknown parameters in the return and volatility equations can be identified using several possible auxiliary models, including those that explicitly capture the relationship between volatility, realized variance, and bipower variation, or various combinations of these components.  However, for reasons of parsimony, we only specify an auxiliary
model for returns and volatility.  In particular, we follow \cite{frazier+mmm19} and use a TARCH-T auxiliary model (threshold GARCH auxiliary model with student-t errors).  
\begin{flalign*}
r_{t}&=\sigma_{t} \varepsilon_{t},\quad \varepsilon_{t} \stackrel{iid}{\sim} t(\nu)\\
\sigma_{t}^{2}&=\gamma_{1}+\gamma_{2} (r_{t-1}-\gamma_0)^2+\gamma_3 \1\left[r_{t-1}<0\right](r_{t-1}-\gamma_0)^2+\gamma_{4} \sigma_{t-1}^{2}
\end{flalign*}with $\gamma=(\gamma_1,\dots,\gamma_4,\nu)'$, and where $\1[A]$ denotes the indicator function on the set $A$. 
The summary statistics $S_2$ are obtained by evaluating the score vector for the auxiliary model, evaluated at the quasi maximum likelihood estimate (QMLE) for the observed data. 

\subsubsection{Simulation of the Model}
Simulating data from the model can be done using
an Euler discretization scheme with step size $1/I$ and $I$ large. 
Write $p_{t,i/I}=p_{t+i/I}$, and define $V_{t,i/I}$ and $\delta_{t,i/I}$ similarly.  Define $\Delta N_{t,i / I}=N_{t,(i+1)/I}-N_{t,i/I}$. 
We discretize the system and generate data recursively: for each day $t$, and each intra-day time $i=1,\dots,I$, simulate data according to
\begin{flalign*}
p_{t,(i+1) / I}&=p_{t, i / I}+\mu \frac{1}{I}+\exp \left(V_{t, i / I} / 2\right) \epsilon_{t, i}^{p} \frac{1}{\sqrt{I}}+Z_{t, i} \Delta N_{t, i / I}\\
V_{t,(i+1) / I}&=V_{t, i / I}+\kappa\left(\alpha-V_{t, i / I}\right) \frac{1}{I}+\frac{\sigma_{v}}{\sqrt{I}}\left(\rho \epsilon_{t, i}^{p}+\sqrt{1-\rho^{2}} \epsilon_{t, i}^{v}\right)
\end{flalign*}where $\epsilon^p,\epsilon^v$ are bivariate standard normal random variables, $Z_{t,i}\stackrel{iid}{\sim}N(\mu_z,\sigma_z^2)$ and $\Delta N_{t,(i+1)/I}$ is $1$ with probability $I^{-1} \delta_{t,(i+1)/I}$, 
$$\delta_{t,(i+1) /I}=\delta_{t,i/I}+(d+\beta \delta_{t,i / I})I^{-1}+\tau \Delta N_{t,i/I},$$ 
and
zero otherwise.  
Given a trajectory $p_{t,i/I}$ define $r_{t,i/I}=p_{t,i / I}-p_{t,(i-1) / I}$, $i=1,\dots, I$.  Then by downsampling
these values to a sequence of $M$ equally spaced values (assuming $I$ is an integer multiple of $M$) 
we obtain $r_{t,i}$, $i=1,\dots, M$, corresponding to the 
$M$ intraday return on day $t$ for the observed data. 

\subsection{Real data analysis}

We consider daily returns data on the S\&P500 index from
26 February 2010 to 7 February 2017.   There are 1750
daily observations, and the most recent 250 observations are reserved for out-of-sample predictive assessments 
using a related, but more parsimonious, model described in the next subsection.  
These data were also used in \cite{frazier+mmm19}.  
Uniform priors are used for each parameter with lower and upper bounds for each parameter given in Table \ref{tab:one}. 
\begin{table}[h!]
	\centering
	\centering
	\caption{{Lower and upper bounds for uniform priors.}}
	\begin{tabular}{lrrrrrrrrrrr}
		& Parameter&  $\mu_p$&$\kappa$&$\alpha$&$\sigma_v$&$\rho$ &$\mu_z$&$\sigma_z$&$d$ & $\tau$ & $\beta$ \\ \hline\hline
		& Lower& -0.1 & 0.05 &-1.0 & 0.001 &  -0.70 &-1.0 &.0.0 & 0.01 & 0.001 & 0.5 \\
		& Upper & 0.1 & 0.50 &3.0  &1.99 &0.0          &1.0 & 3.0 &  0.2  & 0.2 & 1-$\tau$ \\
	\end{tabular}%
	\label{tab:one}%
\end{table}

\begin{figure}[H]
\centering
\includegraphics[width=150mm]{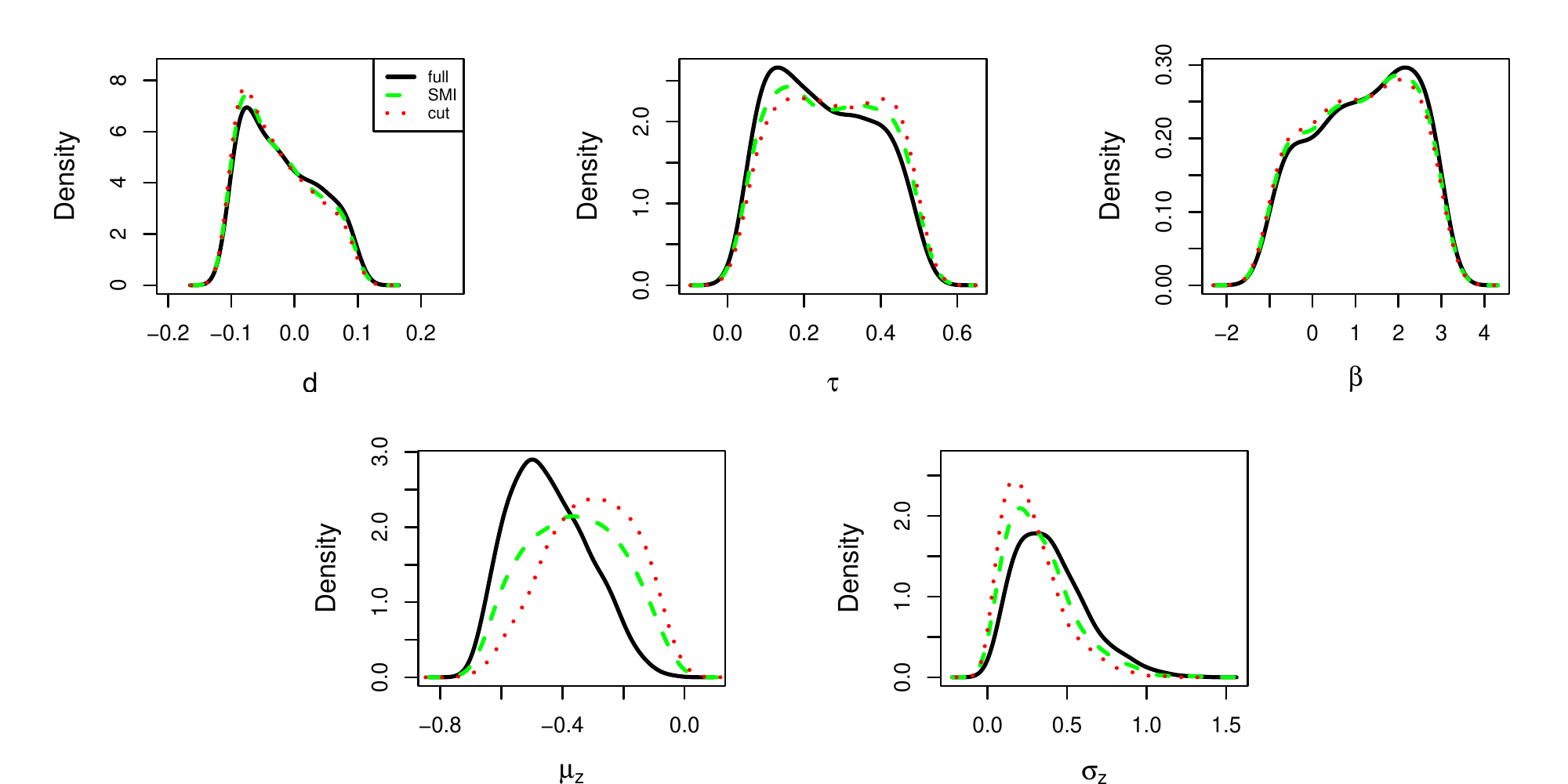}
\caption{\label{smifig}Marginal densities for jump parameters in asset pricing model with jumps for the full, cut and semi-modular posterior densities}
\end{figure}

Figure \ref{smifig} shows the marginal posterior distributions for the parameters $\varphi$ in the jump process, and the corresponding ``cut'' and semi-modular
marginal posterior densities.  The influence parameter $\gamma$ in the semi-modular approach is chosen as
described in Section 4.1, which results in the value $\gamma=0.41$.  The parameters $\mu_z$ and $\sigma_z$ in the jump model, which represent the (average) magnitude and variability of the jumps respectively, are estimated quite differently in the full and
cut posterior distributions;  the interpretation is that cutting feedback suggests that the magnitude of the jumps in daily returns are smaller (i.e., closer to zero) than under the full posterior, and that the variability of the jump size is smaller than under the full posteriors.

The marginal posterior distributions were estimated by fitting a Gaussian mixture model to
50,000 simulations from the prior for parameters and summary statistics.  Similar to the first example, variables
are transformed to be marginally univariate normal before fitting the mixture, with transformation back
to the original scale for estimation of the posterior densities.  The 
\texttt{mclust} package \citep{scrucca+fme16} was used to choose the number of mixture components up to a maximum of 10 by BIC, 
considering different covariance structures for the components.  The final model had 8 mixture
components, with distinct and unrestricted component covariance matrices.  

\begin{figure}[H]
\begin{center}
\includegraphics[width=150mm]{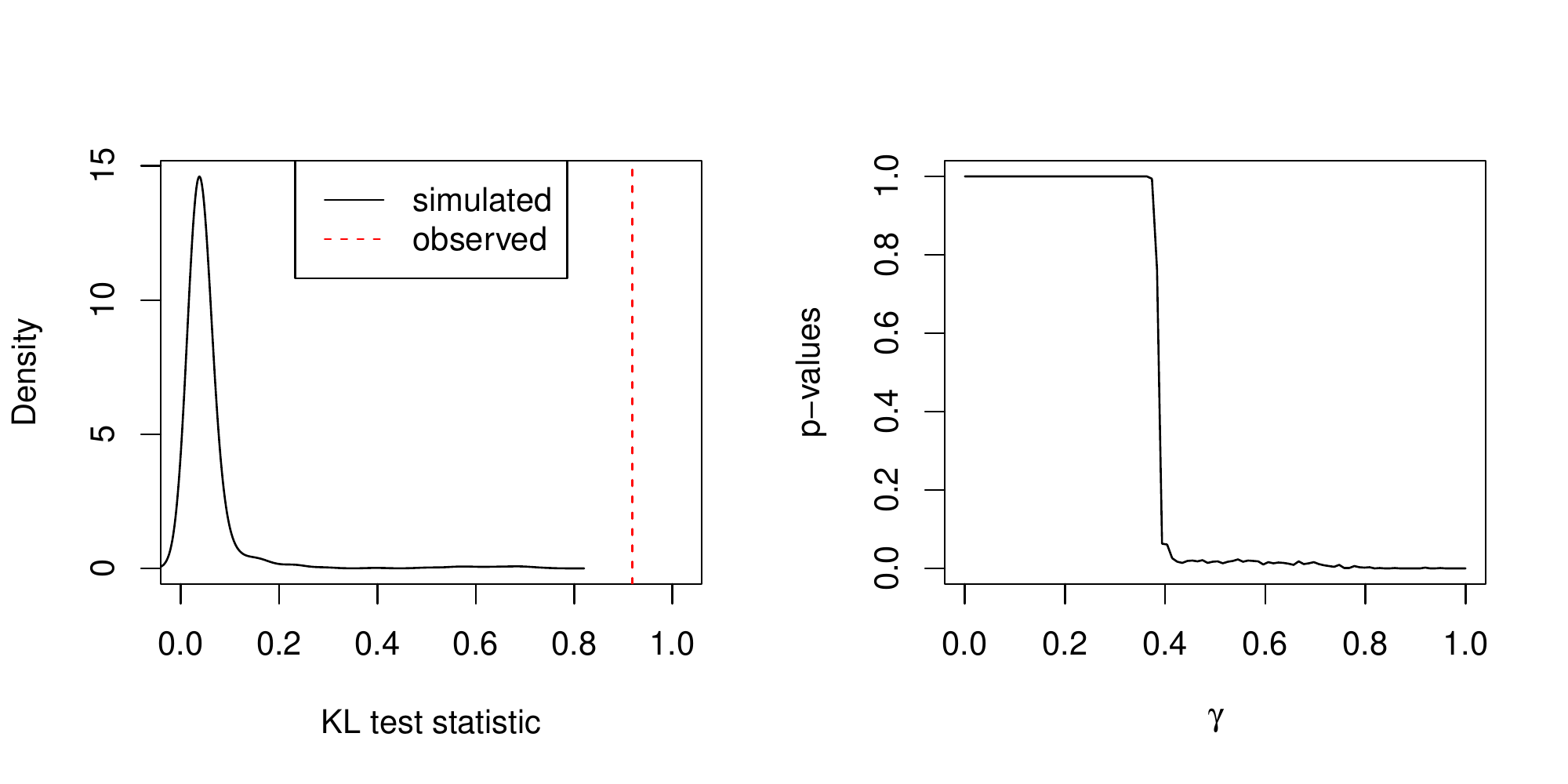} 
\end{center}
\caption{\label{diag} Observed statistic for conflict check $\widetilde{G}(S_{\text{obs},2}|S_{\text{obs},1})$ (left), shown
by the red line, within the reference distribution for the check and the tail probability $p(\gamma)$ as $\gamma$
varies for the semi-modular approach (right), for the asset pricing example.}
\end{figure}

Figure \ref{diag} (left) shows the location of the observed checking
statistic $\widetilde{G}(S_{\text{obs},2}|S_{\text{obs},1})$ (shown by the red line) within its corresponding 
reference distribution, demonstrating that the 
tail probability (\ref{tail}) is very small, and hence that the cut and full posterior distributions
are surprisingly different under the reference distribution for the check, supporting the decision to cut.  
Figure \ref{diag} (right) shows
how the tail probability (\ref{pgamma}) varies with $\gamma$ in the semi-modular approach.  

\begin{figure}[H]
\centerline{\includegraphics[width=150mm]{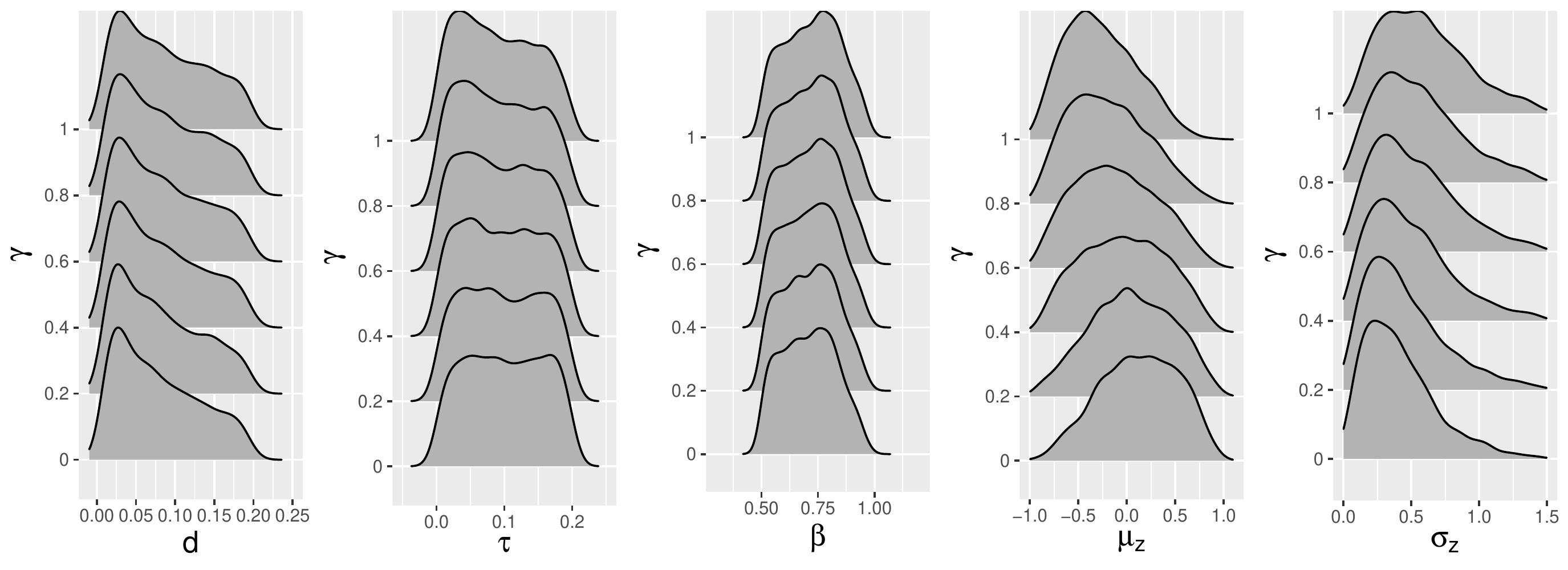}}
\caption{\label{smigamma}Marginal densities for jump parameters in asset pricing model for the semi-modular posterior densities as $\gamma$ changes.}
\end{figure}

Figure \ref{smigamma} shows
how the semi-modular inference for the various parameters in the jump process change with $\gamma$.  Our chosen
value of $\gamma$ was $0.41$ here, and the Figure confirms that for larger values of $\gamma$ the posterior
inference is very different for the corresponding semi-modular and full posterior, particularly for $\mu_z$ and $\sigma_z$.  


While the continuous-time model in equations \eqref{returns}-\eqref{jump} yields parameters that have meaningful structural interpretations, it is well-known that forecasts for returns obtained from continuous-time models  are often outperformed by more parsimonious discrete-time models. Furthermore, the use of continuous-time models in forecasting is hindered by their computational complexity.  Producing forecast densities from continuous-time models requires sequentially approximating the transition density for the states in continuous time, and the conditional density for the observable variables given the states. This approximation is carried out by discretization of the process over a very fine grid, and then simulating forward the process to accurately capture the dynamics. This procedure must then be repeated each time one wishes to produce a forecast. We also note that such an approximation yields a discretization error at each step, which can add additional noise to the resulting forecasts. 

Consequently, in order to produce accurate forecasts, we follow \cite{frazier+mmm19} 
and consider a more parsimonious discrete-time
analogue of the continuous-time model, with likelihood-free inference 
using summary statistics employed to produce the posterior for $\theta$, and a particle filter being used 
for estimation of the latent states in the production of the forecasts.  The resulting model has a similar structural interpretation to the continuous-time model in \eqref{returns}-\eqref{hawkes} but is computationally much simpler to simulate, and does not require any discretization to produce predictive densities. 

\subsection{Discrete time model}

With similar notations to the previous subsection, and following \cite{frazier+mmm19}, we consider the following discrete
time model for daily logarithmic returns and bipower variation:
\begin{align}
r_t & = \exp\left(\frac{h_t}{2}\right) \epsilon_t+\Delta N_t Z_t,\;\;\epsilon_t\sim N(0,1),\; Z_t\sim N(\mu_z,\sigma_z^2),  \label{returns-d} \\
\log \text{BV}_t & = \psi_0+\psi_1 h_t+\sigma_{\text{BV}}\zeta_t, \;\; \zeta_t\sim N(0,1),  \label{bipower-d} \\
h_t & = \omega +\rho h_{t-1}+\sigma_h \eta_t, \;\; \eta_t\sim S(\alpha,-1,0,1), \label{volatility-d}
\end{align}
where $S(\alpha,\beta,\mu,\sigma)$ denotes the $\alpha$-stable distribution with stability parameter $\alpha$, skewness $\beta$, location $\mu$ and scale $\sigma$, and
\begin{align}
P\left(\Delta N_t = 1\mid \mathcal{F}_{t-1}\right) & =\delta_{t}=d+\beta \delta_{t-1}+\tau \Delta N_{t-1}, \label{jumps-d}
\end{align}
where $\mathcal{F}_t$ denotes the $\sigma$-field generated by the observations up to time $t$ and  
$\Delta N_t$ plays a similar role to the continuous time jump process $\dt N_t$ in (\ref{jump}).  
The above model has a similar motivation and structure to the continuous-time model, but as discussed earlier it
is more convenient for forecasting.  It includes an additional measurement equation (\ref{bipower-d}) 
which depends on intra-day returns, through $\ln \text{BV}_t$.
We fix the parameter $\psi_0$ to $0$, since this parameter is hard to identify.  
The set of unknown parameters is $\theta=(\varphi^\top,\eta^\top)^\top$, where 
$\varphi=(\mu_z,\sigma_z,d,\beta,\tau)^\top$ are the parameters in the jump module
and $\eta=(\psi_0,\psi_1,\sigma_{\text{BV}},\omega,\rho,\sigma_h)^\top$ are the parameters in the return/volatility module. 
Our prior distributions are the same as those in \cite{frazier+mmm19}, except for the fixed parameter $\psi_0$. 
 
Once again write the summary statistics as $S=(S_1^\top,S_2^\top)^\top$, 
where $S_1$ is informative about $\varphi$.  
 Our summary statistics are related to those used in \cite{frazier+mmm19} for their TARCH-T auxiliary model, 
 which performed best for forecasting in their work.  
These summary statistics were also the motivation for those used in our continuous-time
model, but must be modified here, since
the jumps-variation $\text{JV}_t$ cannot be computed.  
The reason is that the discrete-time model above only generates the summary of intra-day
returns $\log \text{BV}_t$ directly, without generating the intra-day returns themselves which would be needed
to compute the realized volatility.  We consider for $S_1$ the 
summary statistics obtained from the TARCH-T auxiliary model, so that $S_1$ contains $5$ summary statistics.
The jump process appears only in the model for the daily returns (\ref{returns-d}) in the discrete model, 
and so it is sensible to use the TARCH-T auxiliary
model fitted to the daily returns data 
to summarize the information about the jump process.  For $S_2$, we consider summaries
based on $\log \text{BV}_t$.  We consider the mean, variance and skewness of both $\log \text{BV}_t$, 
and $\log \text{BV}_t-\log \text{BV}_{t-1}$, as well as the correlation of $\text{BV}_t$ and $\text{BV}_{t-1}$, so
that $S_2$ contains 7 summary statistics.

\begin{figure}[H]
\centering
\includegraphics[width=150mm]{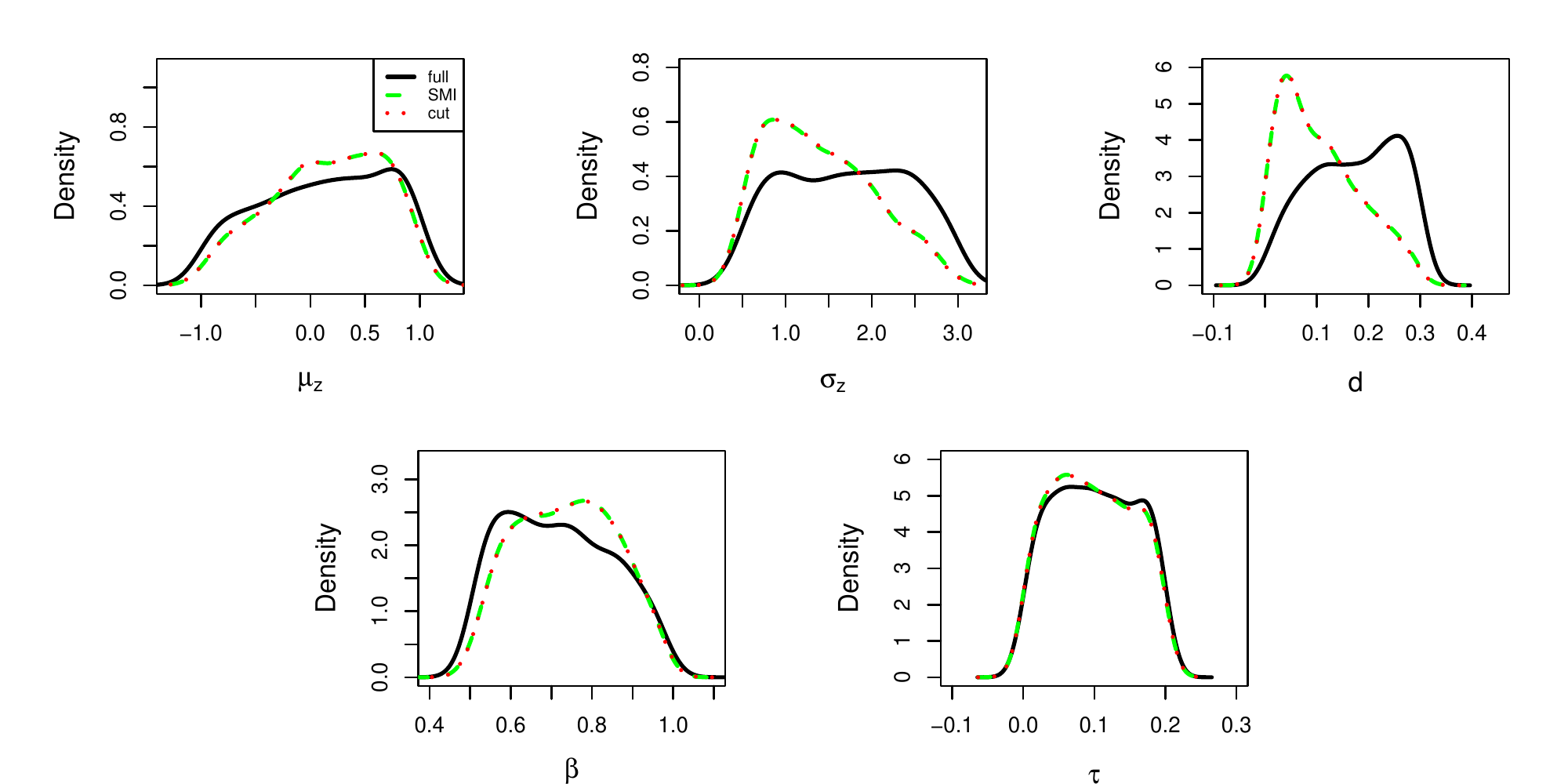}
\caption{\label{smi-forecasting}Marginal densities for jump parameters in discrete time 
asset pricing model with jumps for the full, cut and semi-modular posterior densities.  The cut
and semi-modular posterior densities are similar in all panels.}
\end{figure}

Similar to Figure \ref{smi} for the continuous-time model, 
Figure \ref{smi-forecasting} shows the marginal posterior distributions for the parameters $\varphi$ in the jump process
for the discrete time model, and the corresponding ``cut'' and semi-modular
marginal posterior densities.  Once again, the influence parameter $\gamma$ in the semi-modular approach is chosen as
described in Section 4.1, resulting in $\gamma=0.01$ here so that the SMI posterior is nearly identical to
the cut posterior.  The marginal posterior distributions for $d$ and $\sigma_z$ in 
the cut posterior suggest the presence of fewer jumps
with less variation around the average jump magnitude.
The marginal posterior distributions were estimated by fitting a Gaussian mixture model to
50,000 simulations from the prior for parameters and summary statistics.  Once again, variables
are transformed to be marginally univariate normal before fitting the mixture, and the
\texttt{mclust} package \citep{scrucca+fme16} was used to choose the number of mixture components up to a maximum of 10 by BIC, 
considering different covariance structures for the components.  The final model had 9 mixture
components, with distinct and unrestricted component covariance matrices.  

To assess the forecast performance, we first estimate the posterior distribution of $\theta$ based on the training set
observations.  This estimate is then kept fixed throughout the forecast period, and a bootstrap filter \citep{gordon+ss93} is used
to estimate the latent states $h_t,\Delta N_t$, conditional on the data up to time $t$.  We write this data as
$r_{1:t}=(r_1,\dots, r_t)^\top$, $\text{BV}_{1:t}=(\text{BV}_1,\dots, \text{BV}_t)^\top$.  
Write $y_t=(r_{1:t}^\top,\text{BV}_{1:t}^\top)^\top$. 
We use $5,000$ particles
in the particle filter.  To obtain a one-step ahead forecast at time $T$, write $h_{T,p}^s$, $\Delta N_{T,p}^s$ 
for the $p$th particle values respectively for $h_{T}$, $\Delta N_T$ for posterior parameter sample $\theta^s$, $s=1,\dots, \mathcal{S}$, $p=1\dots, P$.  Write $\delta_{T,p}^s$ for the corresponding values of $\delta_t$.  We use $\mathcal{S}=1,000$.  The one-step ahead forecast density is approximated
by 
\begin{align*}
  \widehat{p}(y_{T+1}|y_{1:T}) \approx \frac{1}{\mathcal{S} \times P}
   \sum_{s=1}^{\mathcal{S}}\sum_{p=1}^P p(y_{T+1}|h_{T+1}=\widetilde{h_{T+1,p}^s}, \Delta N_{T+1}=\widetilde{\Delta N_{T+1,p}^s},\theta),
\end{align*}
where $\widetilde{h_{T+1,p}^s}$ and $\widetilde{\Delta N_{T+1,p}^s}$ are obtained by simulating from (\ref{volatility-d})-(\ref{jumps-d}) with $h_{T}=h_{T,p}^s$ and $\delta_T=\delta_{T,p}^s$,  
and $p(y_t|h_t,\Delta N_t,\theta)$ is defined from equation (\ref{returns-d}).  Similar to \cite{frazier+mmm19}, 
out-of-sample predictive performance for the one-step ahead forecasts are assessed by average predictive log score, 
quadratic score, and continuous ranked probability score, with results shown in Table \ref{scores}.  
\begin{table}
\renewcommand\arraystretch{1.2}
\begin{center}
\caption{\label{scores}Logarithmic (LS), quadratic (QS) and continuous rank probability score (CRPS) for out of sample
	forecast assessment for outcomes $r_t$ and $\log \text{BV}_t$ for the discrete time model for semi-modular inference with $\gamma=0.2$, $0.4$, $0.6$ and $0.8$.  $\gamma=0.01$ is the value of the SMI influence parameter chosen by the 
	conflict checking method, giving similar predictive scores to the cut posterior with $\gamma=0$.  $\gamma=1$ is the full posterior.  
	Scores are oriented so that larger values represent better forecasting performance. 
	}
\begin{tabular}{ccccccc}
  \hline\hline
          & $\gamma=0$ & $\gamma=0.2$ & $\gamma=0.4$ & $\gamma=0.6$ & $\gamma=0.8$ & 
$\gamma=1.0$  \\
& \multicolumn{6}{l}{Outcome $r_t$}\\ \cline{2-7}
LS & -0.670 & -0.673 & -0.677 & -0.682 & -0.686 & -0.690 \\
QS & 0.612 & 0.615 & 0.615 & 0.616 & 0.616 & 0.617 \\
CRPS & -0.259 & -0.259 & -0.260 & -0.260 & -0.260 & -0.260 \\
& \multicolumn{6}{l}{Outcome $\log \text{BV}_t$}\\ \cline{2-7}
LS & -3.894 & -3.877 & -3.914 & -3.909 & -3.926 & -3.924 \\
QS & -0.231 & -0.231 & -0.232 & -0.228 & -0.228 & -0.228 \\
CRPS & -1.801 & -1.806 & -1.819 & -1.823 & -1.833 & -1.840 
\end{tabular}
\end{center}
\end{table}
Although the largest logarithmic and CRPS score values occur for
small values for $\gamma$ for both outcomes, the differences in
forecasting performance between methods
are minor in any practical sense.   

\section{Discussion}

Cutting feedback methods are useful in applications involving multi-modular models, where they can be
used both as a diagnostic
for understanding misspecification and posterior sensitivity as well as an alternative to using the full posterior for 
predictive inference when the development of an alternative model is infeasible.
As far as we are aware, the use of cutting feedback methods has so far been restricted in the literature to
models with tractable likelihood.  The extension to the intractable likelihood setting discussed here can
be useful when it is desired to restrict the information used for inference about a subset
of the parameters to that obtained from a subset of the summary statistics only.  Our proposed Gaussian mixture
model approach to estimation of the posterior distribution makes the cutting feedback computations easy
to perform, and facilitates model checks which can help guide the decision of whether or not to cut, as well
as a semi-modular inference extension where feedback is partially cut.  

The use of cutting feedback methods in likelihood-free inference is particularly helpful since model misspecification 
is known to negatively impact common likelihood-free inference procedures.  In the case of ABC, 
\cite{frazier2020model} demonstrate that if the model is misspecified, then the ABC posterior does not produce 
valid inferences and can be ill-behaved.  \cite{frazier2021synthetic} show that similar problems occur 
for Bayesian synthetic likelihood approaches.  Hence a benefit of cutting feedback is 
that it can be used to hedge against the potential consequences of using misspecified models in 
likelihood-free inference.  Recently, \cite{pacchiardi+d21} consider a 
generalized Bayesian approach to likelihood-free inference 
based on scoring rules which can deal with misspecification.

There has been renewed interest recently in the idea of robustifying Bayesian inference by
conditioning on an
insufficient data summary to exclude information \citep{li+nfs15,lewis+ml21}.  This is interesting regardless
of whether the likelihood is tractable or not, and the consideration of complex data
summaries for conditioning 
leads to possible applications of likelihood-free inference methods in models with tractable likelihood when
misspecification is a concern.  
Another recent work which is relevant to the likelihood-free inference literature is \cite{miller+d19}, where
the authors make a connection between inference for a ``coarsened'' version of the data, which is
reminiscent of ABC methods, and the use of power posterior distributions.  
In traditional ABC methods the ABC kernel can be interpreted in terms of an allowance
for model misspecification \citep{wilkinson13}, and an interesting avenue for future research might be to pursue
ABC approaches to the semi-modular inference framework making use of this interpretation as a way
of partially cutting feedback.  This seems related to the kernel-smoothing $\delta$-SMI approach discussed in 
\cite{nicholls2022valid}.

\section*{Acknowledgements}

DTF is supported by the Australian Research Council. CD is supported by the Australian
Research Council through the Future Fellowship scheme (FT210100260).  SAS is supported by the Australian
Research Council through the Discovery Project scheme (FT170106079) and the ARC Centre of Excellence
in Mathematical and Statistical Frontiers (ACMS; CE140100049).  

\bibliographystyle{chicago}
\bibliography{references}

\end{document}